\documentclass[12pt,a4paper]{article}
\usepackage{amssymb,amsmath}
\usepackage{amsfonts}
\usepackage{a4}
\usepackage{enumerate,verbatim,tikz}

\newtheorem{theorem}{Theorem}

\newtheorem{cor}{Corollary}

\begin{document}
 \baselineskip18pt
	
	\title{\bf Skew  constacyclic codes over a non-chain ring $\mathbb{F}_{q}[u,v]/\langle f(u),g(v), uv-vu\rangle$}
	\author{Swati Bhardwaj{\footnote {E-mail: swatibhardwaj2296@gmail.com}} ~ and ~ Madhu Raka{\footnote {Corresponding author, e-mail: mraka@pu.ac.in}}
		\\ \small{\em Centre for Advanced Study in Mathematics}\\
		\small{\em Panjab University, Chandigarh-160014, INDIA}\\
		\date{}}
	\maketitle
\maketitle
{\abstract{Let $f(u)$ and $g(v)$ be two polynomials of degree $k$ and $\ell$ respectively, not both linear, which split into distinct linear factors over $\mathbb{F}_{q}$. Let $\mathcal{R}=\mathbb{F}_{q}[u,v]/\langle f(u),g(v),\\uv-vu\rangle$ be a finite commutative non-chain ring. In this paper, we study $\psi$-skew cyclic and $\theta_t$-skew constacyclic codes over the ring $\mathcal{R}$ where $\psi$ and $\theta_t$ are two automorphisms defined on  $\mathcal{R}$.
\vspace{2mm}
		}
	\section{ Introduction}

Cyclic codes over finite fields have been studied since 1960's because of their algebraic structures as ideals in certain commutative rings. Interest in
codes over finite rings increased substantially after a break-through work by Hammons et al.  in 1994.
In 2007, Boucher et al. \cite{Bou2007} generalized the concept of cyclic code  over a non-commutative ring, namely skew polynomial ring $\mathbb{F}_{q}[x;\theta]$, where $\mathbb{F}_{q}$ is a field with $q$ elements and $\theta$ is an automorphism of $\mathbb{F}_{q}$.
In the polynomial ring $\mathbb{F}_{q}[x;\theta]$, addition is defined as the usual one of polynomials and the multiplication is defined by the rule
$ax^i*bx^j=a\theta^i(b)x^{i+j}$ for $a,b \in \mathbb{F}_{q}$.  Boucher and Ulmer \cite{BU2009} constructed some $\theta$-cyclic codes  called skew cyclic codes with Hamming distance larger than that of  previously known linear codes with the same parameters.  Siap et al. \cite{Siap2011}  investigated structural properties of skew cyclic codes of arbitrary length. \vspace{2mm}

 After the first phase of study on skew cyclic codes over fields, the focus of attention moved to skew cyclic codes over rings.
 Abualrub et.al \cite{Abu2012} studied skew cyclic codes over
$\mathbb{F}_{2}+v\mathbb{F}_{2}$, where $v^2=v$ and   the automorphism $\theta$ was taken as  $\theta : v\rightarrow v+1$. Li Jin \cite{Jin} studied skew cyclic codes over
$\mathbb{F}_{p}+v\mathbb{F}_{p}$, where $v^2=1$ with   the automorphism $\theta$  taken as  $\theta : a+bv\rightarrow a-bv$.
  In 2014, Gursoy et al. \cite{Gur2014} determined generator polynomials and  found idempotent generators of skew cyclic codes over $\mathbb{F}_{q}+v\mathbb{F}_{q}$, where $v^2=v$ and   the automorphism $\theta$ was defined as $\theta_t : a+bv \rightarrow a^{p^t}+b^{p^t}v$. Minjia Shi et al. \cite{SHI2015} studied  $\theta_t$-skew-cyclic codes over $\mathbb{F}_{q}+v\mathbb{F}_{q}+v^2\mathbb{F}_{q},$ where $ v^3=v$. Later Minjia Shi et al. \cite{SHI2017} extended these results to skew cyclic codes over $\mathbb{F}_{q}+v\mathbb{F}_{q}+ \cdots + v^{m-1}\mathbb{F}_{q}$, where $v^m=v$. Gao et al. \cite{Gao} studied skew constacyclic codes over $\mathbb{F}_{q}+v\mathbb{F}_{q}$, where $v^2=v$. \vspace{2mm}

Recently people have started studying   skew cyclic codes over finite commutative non-chain rings having 2 or more variables. Yao, Shi and Sol$\acute{e}$ \cite{YSS2015} studied skew cyclic codes over $\mathbb{F}_{q}+u\mathbb{F}_{q}+v\mathbb{F}_{q}+uv\mathbb{F}_{q}$, where $u^2=u, v^2=v, uv=vu$ and $q$ is a prime power. Ashraf and Mohammad \cite{AM2018Asian} studied  skew-cyclic codes over $\mathbb{F}_{q}+u\mathbb{F}_{q}+v\mathbb{F}_{q},$ where $u^2=u, v^2=v, uv=vu=0$. Islam and Prakash \cite{IPIJICOT2018} studied skew cyclic and skew constacyclic codes
over $\mathbb{F}_{q}+u\mathbb{F}_{q}+v\mathbb{F}_{q}+uv\mathbb{F}_{q}$, where $u^2=u, v^2=v$ and $uv=vu$. Islam, Verma and Prakash \cite{IVPIJICOT} studied skew constacyclic codes of arbitrary length over $\mathbb{F}_{p^m}[v,w]/ < v^{2}-1,w^{2}-1,vw-wv>$.  In all these papers $\theta$ was taken as $\theta_t : a \rightarrow a^{p^t}$ defined on $\mathbb{F}_{q}$.  \vspace{2mm}

In this paper, we study skew cyclic and skew constacyclic codes over a more general ring. Let $f(u)$ and $g(v)$ be two polynomials of degree $k$ and $\ell$ respectively,  which split into distinct linear factors over $\mathbb{F}_{q}$.  We assume that at least one of $k$ and $\ell$ is $\geq 2$. Let $\mathcal{R}=\mathbb{F}_{q}[u,v]/\langle f(u),g(v),uv-vu\rangle$ be a finite  non-chain ring. Cyclic codes over this ring $\mathcal{R}$ were discussed in \cite{GR4}. A Gray map is defined from $\mathcal{R}^n \rightarrow \mathbb{F}^{k\ell n}_q$ which preserves  duality. We define two automorphisms $\psi$ and $\theta_t$ on  $\mathcal{R}$ and discuss $\psi$-skew cyclic and $\theta_t$-skew $\alpha$-constacyclic codes over this ring, where $\alpha$ is any unit in $\mathcal{R}$ fixed by the automorphism $\theta_t$, in particular when $\alpha^2=1$.
Some structural properties, specially generator polynomials and idempotent generators for  skew constacyclic codes are determined.  We shall show that a  skew cyclic code  over the ring $\mathcal{R}$ is either  a quasi-cyclic code   or  a cyclic code over $\mathcal{R}$. Further
we shall show that Gray image of  a $\theta_t$-skew $\alpha$-constacyclic code of length $n$ over $\mathcal{R}$  is a $\theta_t$-skew $\alpha$-quasi-twisted code of length $k\ell n$ over $\mathbb{F}_{q}$  of index $k\ell$.  Some examples are also given to illustrate the theory. \vspace{2mm}

 In \cite{RKG}, Raka et al. had discussed $\alpha$-constacyclic codes over the ring $\mathbb{F}_{p}[u]/\langle u^4-u\rangle$, $p\equiv 1 ({\rm mod}~3)$ for a specific unit $\alpha=(1-2u^3)$. (Note that the  unit $\alpha$ here satisfies $\alpha^2=1$ in the ring $\mathbb{F}_{p}[u]/\langle u^4-u\rangle$).
 On taking $\theta_t$ as identity automorphism, the results on $\theta_t$-skew $\alpha$-constacyclic codes (Section 4) give  the corresponding results for $\alpha$-constacyclic code over $\mathcal{R}$ which  generalize the results of \cite{RKG}.  \vspace{2mm}

  The results of this paper can easily be extended over the more general ring\\
$\mathbb{F}_{q}[u_1,u_2,\cdots, u_r]/\langle f_1(u_1),f_2(u_2),\cdots f_r(u_r), u_iu_j-u_ju_i\rangle$
 where polynomials $f_i(u_i)$,   $1\le i\leq r$,  split into distinct linear factors over $\mathbb{F}_{q}$.\vspace{2mm}
	
The paper is organized as follows:   In Section 2, we recall the ring $\mathcal{R} = \mathbb{F}_{q}[u,v]/\langle f(u),g(v),uv-vu\rangle$ and  the Gray map $\Phi$ : $\mathcal{R}^n \rightarrow \mathbb{F}^{k\ell n}_q$. In Section 3, we define two automorphisms $\psi$ and $\theta_t$ on $\mathcal{R}$, while in Sections 3.1, we discuss skew cyclic codes over $\mathcal{R}$ with respect to $\psi$. In Section 4, we study skew $\alpha$-constacyclic codes over the ring $\mathcal{R}$  with respect to the automorphism $\theta_t$.

\section{ The ring $\mathcal{R}$ and the Gray map}
\subsection{The ring $\mathcal{R}$ }
		Let $q$ be a prime power, $q = p^{s}$. Throughout the paper, $\mathcal{R}$ denotes the commutative ring $\mathbb{F}_{q}[u,v]/\langle f(u),g(v), uv-vu\rangle$, where $f(u)$ and $g(v)$ are polynomials of degree $k$ and $\ell$ respectively, which split into distinct linear factors over $\mathbb{F}_q$. We assume that at least one of $k$ and  $\ell$ is $\geq 2$, otherwise $\mathcal{R}\simeq \mathbb{F}_q$. If $\ell=1$ or $k=1$, then the ring $\mathcal{R} = \mathbb{F}_{q}[u,v]/\langle f(u),g(v), uv-vu\rangle$ is isomorphic to $\mathbb{F}_{q}[u]/\langle f(u)\rangle$ or $\mathbb{F}_{q}[v]/\langle g(v)\rangle$.   Duadic and triadic cyclic codes, duadic negacyclic codes over $\mathbb{F}_{q}[u]/\langle f(u)\rangle$ have  been discussed by Goyal and Raka in \cite{GR2, GR3}. Further in \cite{GR4,GR5}, Goyal and Raka have discussed  polyadic cyclic codes and polyadic constacyclic codes over $\mathcal{R}=\mathbb{F}_{q}[u,v]/\langle f(u),g(v), uv-vu\rangle$.\vspace{2mm}\\
 Let $f(u)=(u-{\alpha}_1)(u-{\alpha}_2)...(u-{\alpha}_k)$, with $\alpha_i \in \mathbb{F}_q$, $\alpha_i \neq \alpha_j$ and $g(v)=(v-{\beta}_1)(v-{\beta}_2)...(v-{\beta}_\ell)$, with $\beta_i \in \mathbb{F}_q$, $\beta_i \neq \beta_j$. $\mathcal{R}$ is a  non chain ring of size ${q}^{k\ell}$ and characteristic $p$. \vspace{2mm}\\
		For $k \geq 2$ and $\ell \geq 2$, let $\epsilon_i$, $1 \leq i \leq k$ and $\gamma_j$, $1\leq j\leq \ell$,  be elements of the ring $\mathcal{R}$ given by\vspace{2mm}
		\begin{equation} \begin{array}{ll}
			\epsilon_i=\epsilon_i(u)= \frac{(u-\alpha_1)(u-\alpha_2)\cdots(u-\alpha_{i-1})(u-\alpha_{i+1})\cdots(u-\alpha_k)}{(\alpha_i-\alpha_1)(\alpha_i-\alpha_2)\cdots(\alpha_i-\alpha_{i-1})
(\alpha_i-\alpha_{i+1})\cdots(\alpha_i-\alpha_k)} {~~\rm and}\vspace{4mm}\\
		\gamma_j=\gamma_j(v)= \frac{(v-\beta_1)(v-\beta_2)\cdots(v-\beta_{j-1})(v-\beta_{j+1})\cdots(v-\beta_\ell)}{(\beta_j-\beta_1)(\beta_j-\beta_2)\cdots(\beta_j-\beta_{j-1})
(\beta_j-\beta_{j+1})\cdots(\beta_j-\beta_\ell)}.
		
		\end{array}\vspace{2mm}\end{equation}
If $k \leq 1$,  we define $\epsilon_i=1$ and if   $\ell \leq 1$, we take  $\gamma_j=1$.\vspace{2mm}\\
We note that $\epsilon_i^2=\epsilon_i, ~\epsilon_i\epsilon_r=0~ {\rm ~for~} 1\leq i, r \leq k,~ i \neq r ~{\rm~ and ~}  \sum_{i} \epsilon_i=1$ modulo $f(u)$;  $\gamma_j^2=\gamma_j, ~\gamma_j\gamma_s=0~ {\rm ~for~} 1\leq j, ~s \leq \ell, ~j \neq s ~{\rm~ and ~}  \sum_{j} \gamma_j=1$  modulo $g(v)$ in $\mathcal{R}$.\vspace{2mm}

\noindent	 For $ i= 1,2,\cdots,k, j=1,2,...,\ell $, define $\eta_{ij}$ as follows
		\begin{equation}\eta_{ij}=\eta_{ij}(u,v)= \epsilon_i(u) \gamma_j(v). \end{equation}

\noindent{\bf  Lemma ~1:} We have $\eta_{ij}^2=\eta_{ij}, ~\eta_{ij}\eta_{rs}=0~ {\rm ~for~} 1\leq i, r \leq k, 1\leq j, s \leq \ell, (i,j) \neq (r,s) ~{\rm~ and ~}  \sum_{i,j} \eta_{ij}=1$ in $\mathcal{R}$, i.e., $\eta_{ij}$'s are primitive orthogonal idempotents of the ring $\mathcal{R}$.\vspace{2mm}
		
\noindent This is Lemma 2 of \cite{GR4}.\vspace{2mm}

\noindent The decomposition theorem of ring theory tells us that $\mathcal{R}= \underset{i,j}{\bigoplus}~\eta_{ij}\mathcal{R}$. \vspace{2mm}

\noindent For a linear code $\mathcal{C }$ of length $n$ over the ring $\mathcal{R}$, let for each pair $(i,j), 1 \leq i \leq k, 1 \leq j \leq \ell$, \vspace{2mm}

	$\mathcal{C }_{ij}= \{ x_{ij}\in \mathbb{F}_{q}^n : \exists ~x_{rs}  \in \mathbb{F}_{q}^n, (r,s) \neq (i,j),  {\rm ~such ~that~} \underset{r,s}{\sum}~\eta_{rs}x_{rs} \in \mathcal{C }\}.$\vspace{2mm}\\
		Then $\mathcal{C}_{ij}$ are linear codes of length $n$ over $\mathbb{F}_{q}$,  $\mathcal{C}=\underset{i,j}{\bigoplus}~\eta_{ij}\mathcal{C}_{ij}
		$ and $|\mathcal{C }|= \underset{i,j} {\prod}|\mathcal{C }_{ij}|$.
			
\begin{theorem} Let $\mathcal{C}=\underset{i,j}{\bigoplus}~\eta_{ij}\mathcal{C}_{ij}
		$ be a linear code of length $n$ over $\mathcal{R}$. Then \vspace{2mm}
			
	\noindent{\rm(i)}~~~ $ \mathcal{C}^\perp=\underset{i,j}{\bigoplus}~\eta_{ij}\mathcal{C}_{ij}^\perp,$ \vspace{2mm}

	\noindent {\rm(ii)} ~ $ \mathcal{C}$ is self-dual if and only if 	$\mathcal{C}_{ij}$ are self-dual,\vspace{2mm}

\noindent {\rm(iii)} ~ $|\mathcal{C}^\perp|= \underset{i,j} {\prod}~|\mathcal{C}^\perp_{ij}|$.
\end{theorem}
\noindent {\bf Proof:} Let $a=(a_0,a_1,\cdots,a_{n-1})\in \mathcal{C}^\perp$. This gives $a\cdot b=0$ for all $b=(b_0,b_1,\cdots,b_{n-1})\in \mathcal{C}$. Let $a_r= \underset{i,j}{\sum}~\eta_{ij}a_{ijr}$  and $b_r= \underset{i,j}{\sum}~\eta_{ij}b_{ijr}$ for $0\leq r\leq n-1$ where $a_{ijr}, ~b_{ijr}\in \mathbb{F}_q$. Take $a_{ij}=(a_{ij0},a_{ij1},\cdots,a_{ij(n-1)})	$ and $b_{ij}=(b_{ij0},b_{ij1},\cdots,b_{ij(n-1)})$ so that 	 $a_{ij}, b_{ij}\in \mathbb{F}_{q}^n$ and $a=\underset{i,j}{\sum} \eta_{ij}a_{ij}$, $b=\underset{i,j}{\sum} \eta_{ij}b_{ij}$.  As $b\in \mathcal{C}$, we find that  $b_{ij}\in \mathcal{C}_{ij}$. Now $a\cdot b=0$ implies\vspace{2mm}\\
$0=(\sum \eta_{ij}a_{ij0})(\sum \eta_{ij}b_{ij0})+(\sum \eta_{ij}a_{ij1})(\sum \eta_{ij}b_{ij1})+\cdots+(\sum \eta_{ij}a_{ij(n-1)})(\sum \eta_{ij}b_{ij(n-1)})$\vspace{2mm}\\ which gives, using Lemma 1\vspace{2mm}\\$~~~~~~~~~~~~~~\sum \eta_{ij}a_{ij0}b_{ij0}+\sum \eta_{ij}a_{ij1}b_{ij1}+\cdots+\sum \eta_{ij}a_{ij(n-1)} b_{ij(n-1)}=0$\vspace{2mm}\\
i.e. $\sum \eta_{ij}(a_{ij}\cdot b_{ij})=0$. This implies $a_{ij}\cdot b_{ij}=0$ for all $i,j$, where $b_{ij} \in\mathcal{C}_{ij}$. Therefore
$	a_{ij}\in \mathcal{C}_{ij}^\perp$. Hence $ a \in \underset{i,j}{\bigoplus}~\eta_{ij}\mathcal{C}_{ij}^\perp,$ so that $ \mathcal{C}^\perp \subseteq \underset{i,j}{\bigoplus}~\eta_{ij}\mathcal{C}_{ij}^\perp.$ The reverse inclusion can be obtained by reversing the above steps. This proves (i) and (ii), (iii) follow immediately from (i). \hfill $\square$	 
\subsection{ The Gray map}
Every element $r(u,v)$ of the ring $\mathcal{R}=\mathbb{F}_{q}[u,v]/\langle f(u),g(v), uv-vu\rangle$ can be uniquely expressed as $$ r(u,v) = {\displaystyle\sum_{i,j}}~\eta_{ij}a_{ij},$$ where $a_{ij} \in \mathbb{F}_q$ for $1 \leq i \leq k, 1 \leq j \leq \ell$. \vspace{2mm}
 	
\noindent Define a Gray map $\Phi : \mathcal{R}\rightarrow \mathbb{F}_q^{k\ell}$  by
\begin{equation} r(u,v) = \sum_{i,j} \eta_{ij}a_{ij} \longmapsto(a_{11},a_{12},\cdots, a_{1\ell},a_{21},a_{22},\cdots, a_{2\ell},\cdots,a_{k1},a_{k2},\cdots, a_{k\ell}).\end{equation}
	
		This map can be extended from $\mathcal{R}^n$ to  $(\mathbb{F}_q^{k\ell})^n$ component wise i.e. for $r=(r_0,r_1,\cdots,r_{n-1})$, where $r_s =\eta_{11}a_{11}^{(s)}+\eta_{12}a_{12}^{(s)}+\cdots+\eta_{kl} a_{k\ell}^{(s)}~ \in \mathcal{R}$,
 define $\Phi$ as follows
$$\begin{array}{ll} \Phi(r_0,r_1,\cdots,r_{n-1})& = \big( \Phi(r_0),\Phi(r_1),\cdots,\Phi(r_{n-1})\big)\vspace{2mm}\\& =\Big(a_{11}^{(0)},a_{12}^{(0)},\cdots,a_{k\ell}^{(0)}, a_{11}^{(1)},a_{12}^{(1)},\cdots, a_{k\ell}^{(1)}, a_{11}^{(n-1)},\cdots, a_{k\ell}^{(n-1)}\Big).\end{array}$$

\noindent Let the Gray weight of an element $r \in \mathcal{R}$ be $w_{G}(r) =w_H(\Phi(r))$, the Hamming weight of $\Phi(r)$. The Gray weight of  a codeword
		$c=(c_0,c_1,\cdots,c_{n-1})$ $\in \mathcal{R}^n$ is defined as $w_{G}(c)=\sum_{i=0}^{n-1}w_{G}(c_i)=\sum_{i=0}^{n-1}w_H(\Phi(c_i))=w_H(\Phi(c))$. For any two elements $c_1, c_2 \in \mathcal{R}^n$, the Gray distance $d_{G}$ is given by $d_{G}(c_1,c_2)=w_{G}(c_1-c_2)=w_H(\Phi(c_1)-\Phi(c_2))$. The next theorem is a special case of a result of Goyal and Raka \cite{GR4}.

\begin{theorem} The Gray map $\Phi$ is an  $\mathbb{F}_q$ - linear, one to one and onto map. It is also distance preserving map from ($\mathcal{R}^n$, Gray distance $d_{G}$) to ($\mathbb{F}_q^{k\ell n}$, Hamming distance $d_H$). Further $\Phi(\mathcal{C}^{\perp})=(\Phi(\mathcal{C}))^{\perp}$ for any linear code   $\mathcal{C}$  over $\mathcal{R}$. \end{theorem}

Sometimes it is more convenient to use a permuted version of the Gray map $\Phi$ on $\mathcal{R}^n$. For $r=(r_0,r_1,\cdots,r_{n-1})$, where
$r_s=\eta_{11}a_{11}^{(s)}+\eta_{12}a_{12}^{(s)}+\cdots+\eta_{kl} a_{k\ell}^{(s)}~$, define
$\Phi_\pi :\mathcal{R}^n\rightarrow(\mathbb{F}_q^{k\ell})^n$ by

\begin{equation} \begin{array}{l}\Phi_\pi(r_0,r_1,\cdots,r_{n-1}) = \Big(a_{11}^{(0)},a_{11}^{(1)},\cdots,a_{11}^{(n-1)},a_{12}^{(0)},a_{12}^{(1)},\cdots,a_{12}^{(n-1)},\cdots,\\~~~~~~~~~~~~~~~~~~~~~~~~~~~~~~~~~~~~~~~~~~~~~ a_{k\ell}^{(0)}, a_{k\ell}^{(1)},\cdots, a_{k\ell}^{(n-1)}\Big).\end{array}\end{equation}
Clearly  the Gray images  $\Phi(\mathcal{C})$   and $\Phi_\pi(\mathcal{C})$ of a linear code   $\mathcal{C}$  over $\mathcal{R}$ are equivalent codes.

	\section{Skew Cyclic codes over the ring $\mathcal{R}$ }

Let  $\theta$ be an automorphism of $\mathcal{R}$. The map $\theta$ can be extended to $\mathcal{R}^n$  component wise i.e. for $c=(c_0,c_1,\cdots,c_{n-1})$, \begin{equation}\theta(c)=\big(\theta(c_0),\theta(c_1),\cdots,\theta(c_{n-1})\big).\end{equation}

\noindent Let $c=(c_0,c_1,\cdots,c_{n-1})\in \mathcal{R}^n$. The cyclic shift of $\theta(c)$- called $\theta$-cyclic shift or the skew cyclic shift   is defined as
\begin{equation}\sigma_{\theta}(c)=(\theta(c_{n-1}),\theta(c_0),\cdots,\theta(c_{n-2})). \end{equation}
Let $c$ be  divided into $m$ equal parts of length $r$ where $n=mr$, i.e. \vspace{2mm}\\$~~~~~~~~~c=\big(c_{0,0},c_{0,1},\cdots,c_{0,r-1},c_{1,0},\cdots,c_{1,r -1},\cdots,c_{m-1,0},\cdots,c_{m-1,r-1}\big).$\vspace{2mm}\\ Write $c=\big(c^{(0)}|c^{(1)}|\cdots |c^{(m-1)}\big)$.
  The  skew quasi-cyclic shift of $c$ of index $m$ is defined as  \begin{equation}\tau_{\theta,m}(c)=\big(\theta(c^{(m-1)})|\theta(c^{(0)})|\cdots |\theta(c^{(m-2)})\big).\end{equation}

A linear code $\mathcal{C}$ of length $n$ over $\mathcal{R}$ is called a skew cyclic code if $\sigma_{\theta}( \mathcal{C})=\mathcal{C}$ and  a skew quasi-cyclic code of  index $m$ if $\tau_{\theta,m}( \mathcal{C})=\mathcal{C}$.   \vspace{2mm}

The set $\mathcal{R}[x,\theta]= \{a_0+a_1x+a_2x^2+\cdots+a_sx^s : a_i \in \mathcal{R}, ~s \geq 0{\rm ~ integer}\}$, where the variable $x$ is written on the right of the coefficients, forms a ring under usual addition of polynomials and the multiplication is defined  as $ax^i*bx^j=a\theta^i(b)x^{i+j}$ for $a,b \in \mathcal{R}$. The skew polynomial ring  $\mathcal{R}[x,\theta] $ is non-commutative unless $\theta$ is the identity isomorphism. Let $\mathcal{R}_n= \mathcal{R}[x,\theta]/\langle x^n-1\rangle$.  $\mathcal{R}_n$ is a left $\mathcal{R}[x,\theta]$-module with usual addition and left multiplication defined as $ r(x)*(f(x)+\langle x^n-1\rangle)= r(x)*f(x)+\langle x^n-1\rangle$ for $r(x) \in \mathcal{R}[x,\theta]$ and $f(x)+\langle x^n-1\rangle \in \mathcal{R}_n$.
  In polynomial representation, a linear code of length $n$ over  $\mathcal{R}$ is a skew cyclic code if and only if it is a left $\mathcal{R}[x,\theta]$-submodule of  $\mathcal{R}[x,\theta]/\langle x^n-1\rangle$.  \vspace{2mm}

 In polynomial representation, a skew quasi-cyclic code of length $n=mr$ and index $m$ can be viewed as a left $\mathcal{R}[x,\theta]/\langle x^m-1\rangle$-submodule of $\big( \mathcal{R}[x,\theta]/\langle x^m-1\rangle \big)^r$ due to  the one-to-one correspondence  : $\mathcal{R}^{mr}\rightarrow \big(\mathcal{R}[x,\theta]/\langle x^m-1\rangle\big)^r$ given by \vspace{2mm}\\
$\begin{array}{ll}c=&\big(c_{0,0},c_{0,1},\cdots,c_{0,r-1},c_{1,0},\cdots,c_{1,r -1},\cdots,c_{m-1,0},\cdots,c_{m-1,r-1}\big)\\&\rightarrow \big(c_{0,0}+c_{1,0}x+\cdots+c_{m-1,0}x^{m-1}, c_{0,1}+c_{1,1}x+\cdots+c_{m-1,1}x^{m-1},\\& \cdots, c_{0,r-1}+c_{1,r-1}x+\cdots+c_{m-1,r-1}x^{m-1}\big)\end{array}$ \vspace{2mm}

\noindent In this paper, we will consider the following  two  automorphisms  on the ring $\mathcal{R}=\mathbb{F}_{q}[u,v]/\langle f(u),g(v), uv-vu\rangle$.
\begin{enumerate}
\item  Without loss of generality, suppose that $\ell \geq 2$. For an $a \in \mathcal{R}$\vspace{1mm}\\ $a =\sum_{i,j} \eta_{ij}a_{ij}= \sum_{j=1}^{\ell} \eta_{1j}a_{1j} +\sum_{j=1}^{\ell} \eta_{2j}a_{2j}+\cdots +\sum_{j=1}^{\ell} \eta_{kj}a_{kj}$,  define $$ \begin{array}{ll}\psi(a)= & (\eta_{1\ell}a_{11}+\eta_{11}a_{12}\cdots+\eta_{1(\ell-1)}a_{1\ell})+(\eta_{2\ell}a_{21}+\eta_{21}a_{22}\cdots+\eta_{2(\ell-1)}a_{2\ell}) \\& +\cdots  +(\eta_{k\ell}a_{k1}+\eta_{k1}a_{k2}\cdots+\eta_{k(\ell-1)}a_{k\ell})\end{array}.$$
Clearly  the order of $\psi$ is $\ell$.

\item  Let $q=p^s$ and $t$ be an integer $ 1\leq t \leq s$. Define  an automorphism $\theta_t : \mathbb{F}_{q} \rightarrow	\mathbb{F}_{q} $ given by $\theta_t(a)=a^{p^t}$ and extend it to $\theta_{t} : \mathcal{R}\rightarrow \mathcal{R}$ by
  $$\theta_{t}\big(\sum_{i,j} \eta_{ij}a_{ij}\big)= \sum_{i,j} \eta_{ij}a_{ij}^{p^t}.$$
  Note that if $t=s$, $\theta_t$ is the identity map and this automorphism  is irrelevant  if $q$ is a prime.
    		
Clearly  the order of $\theta_{t}$ is $|\theta_{t}|= s/t$ and  the ring $\mathbb{F}_{p^t}[u,v]/\langle f(u),g(v), uv-vu\rangle$ is invariant under $\theta_{t}$.

\end{enumerate}
\subsection{$\psi$-skew Cyclic codes over the ring $\mathcal{R}$}
In this subsection, we discuss skew cyclic codes over $\mathcal{R}$ with respect to automorphisms $\psi$.

\begin{theorem} The center   $Z(\mathcal{R}[x,\psi])$ of $\mathcal{R}[x,\psi]$ is $\mathbb{F}_{q}[x^\ell]$.
\end{theorem}

\noindent {\bf Proof :} Since the order of $\psi$ is $\ell$, for any natural number $i$ and $a\in \mathcal{R}$, we have $x^{\ell i} \ast a=(\psi^\ell)^i(a) x^{\ell i}= a\ast x^{\ell i}$; so $x^{\ell i}$ is in the center of $\mathcal{R}[x,\psi]$. As the fixed ring of $\mathcal{R}$ by $\psi$ is  $\mathbb{F}_{q}$, any $f\in \mathbb{F}_{q}[x^\ell]$ is a central element. Conversely for any $ f\in Z(\mathcal{R}[x,\psi])$ and $a\in \mathcal{R}$, we have $x\ast f=f\ast x$ and $a\ast f=f\ast a$ which implies $f \in \mathbb{F}_{q}[x^\ell]$.\hfill $\square$

\begin{cor} The polynomial $x^n-1$ is in the center   $Z(\mathcal{R}[x,\psi])$ if and only if $\ell$ divides $n$.
\end{cor}

\noindent {\bf Remark 1} If $ \ell| n$, then  $\mathcal{R}_n=\mathcal{R}[x,\psi]/\langle x^n-1\rangle$ is a ring and a skew cyclic code $\mathcal{C}$ of length $n$ over $\mathcal{R}$ is a left ideal in $\mathcal{R}_n$.

\begin{theorem} Let   $\mathcal{C}$  be a skew cyclic code of length $n$. If $g(x)$ is a polynomial in $\mathcal{C}$ of minimal degree and leading coefficient of $g(x)$  is a unit in $\mathcal{R}$, then $\mathcal{C}=\langle g(x)\rangle$ where $g(x)$ is a right divisor of $x^n-1$.
\end{theorem}

\noindent {\bf Proof :} Let $c(x)\in \mathcal{C}$. Write $c(x)=q(x)g(x)+r(x)$ where $q(x), r(x) \in \mathcal{R}[x,\psi]$ and deg $r(x)<$deg $ g(x)$. Since $\mathcal{C}$ is a left $\mathcal{R}[x,\psi]$-submodule, $r(x)=c(x)-q(x)g(x)\in \mathcal{C}$. Therefore we must have $r(x)=0$ and so $\mathcal{C}=\langle g(x)\rangle$. Further if $x^n-1=q(x)g(x)+r(x)$ for some skew polynomials $q(x), r(x) \in \mathcal{R}[x,\psi]$ and deg $r(x)<$deg $g(x)$, then $r(x)=(x^n-1)-q(x)g(x)\in \mathcal{C}$ and so $r(x)=0$. Therefore $g(x)$ is a right divisor of $x^n-1$.\hfill $\square$

\begin{theorem} Let   $\mathcal{C}$ be a skew cyclic code of length $n$ over $\mathcal{R}$ and let $r=\gcd(n,|\psi|)=\gcd (n,\ell)$. If $r=1$, then $\mathcal{C}$ is a cyclic code of length $n$ over $\mathcal{R}$; if $r> 1$ then $\mathcal{C}$ is a quasi-cyclic code of index $n/r$.\end{theorem}

\noindent{\bf Proof:} Let $n=mr$.  Find integers $a$ and $b>0$ such that $ a \ell= r+bn$. (As $\gcd(\ell,n)=r$, there exist integers $a',b'$ such that $a'\ell+b'n=r$. If $b'<0$, we are done. If $b'>0$, find a positive integer $ t$ such that $\ell t-b'>0$. Then $(a'+nt)\ell=r+(\ell t-b')n$.) Let \\$c=\big(c_{0,0},c_{0,1},\cdots,c_{0,r-1},c_{1,0},\cdots,c_{1,r -1},\cdots,c_{m-1,0},\cdots,c_{m-1,r-1}\big)$\vspace{2mm}\\ be a codeword in $\mathcal{C}$ divided into $m$ equal parts of length $r$. \vspace{2mm}\\ Write $c= \big(c^{(0)}|c^{(1)}|\cdots |c^{(m-1)}\big)$. Since $\mathcal{C}$ is a skew cyclic code, $\sigma_{\psi}(c), \sigma_{\psi}^2(c),\cdots, $ $\sigma_{\psi}^r(c),\cdots $ all belong to
$\mathcal{C}$.  Since  $r+bn$ is divisible by $\ell= |\psi|$, we have $$ \begin{array}{rl}~~~\sigma_{\psi}^r(c)&= \big( \psi^r(c^{(m-1)})|\psi^r(c^{(0)})|\cdots |\psi^r(c^{(m-2)})\big)\vspace{2mm}\\
 ~~~\sigma_{\psi}^{2r}(c)&= \big( \psi^{2r}(c^{(m-2)})| \psi^{2r}(c^{(m-1)})|\cdots |\psi^{2r}(c^{(m-3)})\big)\vspace{2mm}\\
 ~~~\sigma_{\psi}^n(c)&= \big( \psi^n(c^{(0)})|\psi^n(c^{(1)})|\cdots |\psi^n(c^{(m-1)})\big)\vspace{2mm}\\
 ~~~\sigma_{\psi}^{bn}(c)&= \big( \psi^{bn}(c^{(0)})|\psi^{bn}(c^{(1)})|\cdots |\psi^{bn}(c^{(m-1)})\big)\vspace{2mm}\\
 \sigma_{\psi}^{r+bn}(c)&= \big( \psi^{r+bn}(c^{(m-1)})|\psi^{r+bn}(c^{(0)})|\cdots |\psi^{r+bn}(c^{(m-2)})\big)\vspace{2mm}\\&=\big( c^{(m-1)}|c^{(0)}|\cdots |c^{(m-2)}\big)=\tau_{\theta,m}(c), \end{array}$$
 where $\theta$ is Identity automorphism.
 Therefore $\tau_{\theta,m}(c)\in \mathcal{C}$. If $r>1$, $\mathcal{C}$ is a quasi-cyclic code of index $m$. If $r=1$, i.e. $m=n$, then
$\mathcal{C}$ is a cyclic code of length $n$ over $\mathcal{R}$. \hfill $\square$\vspace{2mm}

\noindent {\bf Remark 2 } The above result holds for any automorphism $\theta$ on $\mathcal{R}$.\vspace{2mm}

\noindent{\bf Example 1 :} Let $\mathcal{R}=\mathbb{F}_4[u,v]/\langle u(u-1)(u-\alpha), v^2-v, uv-vu\rangle$, where  $\mathbb{F}_4=\mathbb{F}_2[\alpha]$ and  $\alpha^2+\alpha+1=0$. Here $\epsilon_1= \frac{(u-1)(u-\alpha)}{\alpha}$, $\epsilon_2= \frac{u(u-\alpha)}{1-\alpha}$, $\epsilon_3= \frac{u(u-1)}{\alpha(\alpha-1)}$, $\gamma_1=1-v$ and $\gamma_2=v$. We have  $ \gamma_1=1-v =(1\cdot\eta_{11}+0\cdot\eta_{12})+(1\cdot\eta_{21}+0\cdot\eta_{22})+(1\cdot\eta_{31}+0\cdot\eta_{32})$. One finds that  $\psi(1-v)= (0\cdot\eta_{11}+1\cdot\eta_{12})+(0\cdot\eta_{21}+1\cdot\eta_{22})+(0\cdot\eta_{31}+1\cdot \eta_{32})= \gamma_2=v$ and $\psi(v)=1-v$. The order of $\psi$ is $2$. The polynomial $g(x)=x^6+vx^5+x^4+x^3+x^2+(1-v)x+1$ is a right divisor of $x^{12}-1$ over the ring $\mathcal{R}[x,\psi]$, therefore it generates a skew cyclic code of length $12$ over $\mathcal{R}$. By Theorem 5, this code is a quasi-cyclic code of index $6$. \vspace{2mm}

\noindent{\bf Example 2 :} Let $\mathcal{R}=\mathbb{F}_8[u,v]/\langle u(u-1), v(v-1)(v-\beta)(v-\beta^2), uv-vu\rangle$, where  $\mathbb{F}_8=\mathbb{F}_2[\beta]$ and  $\beta^3+\beta+1=0$. Here  $\epsilon_1=1-u$ and $\epsilon_2=u$, $\gamma_1= \frac{(v-1)(v-\beta)(v-\beta^2)}{\beta +1}$, $\gamma_2= \frac{v(v-\beta)(v-\beta^2)}{\beta^2}$, $\gamma_3= \frac{v(v-1)(v-\beta^2)}{\beta}$ and $\gamma_4= \frac{v(v-1)(v-\beta)}{\beta^2+\beta+1}$. We have  $ \gamma_1 =(1\cdot\eta_{11}+0\cdot\eta_{12}+0\cdot\eta_{13}+0\cdot\eta_{14})+(1\cdot\eta_{21}+0\cdot\eta_{22}+0\cdot\eta_{23}+0\cdot\eta_{24})$. One finds that  $\psi(\gamma_1)= \gamma_2$, $\psi(\gamma_2)= \gamma_3$, $\psi(\gamma_3)= \gamma_4$, $\psi(\gamma_4)= \gamma_1$ and  $\psi(\epsilon_i)=\epsilon_i$ for $i=1,2$. The order of $\psi$ is $4$. The polynomial $g(x)=x^4+u(\gamma_1+\gamma_3)x^3+u(\gamma_1+\gamma_3)x+1$ is a right divisor of $x^{8}-1$ over the ring $\mathcal{R}[x,\psi]$, therefore it generates a skew cyclic code of length $8$ over $\mathcal{R}$. By Theorem 5, this code is a quasi-cyclic code of index $2$. \vspace{2mm}

\noindent{\bf Example 3 :} Let $\mathcal{R}=\mathbb{F}_5[u,v]/\langle u(u-1), v(v-1), uv-vu\rangle$ and $n=9$. The polynomial $g(x)=x^6+x^3+1$ generates a skew cyclic code of length $9$ over $\mathcal{R}$. This code is equivalent to a cyclic code of length $9$, by Theorem 5.

\section{$\theta_t$-skew constacyclic codes over the ring $\mathcal{R}$}
In this section we will study $\theta_t$-skew $\alpha$-constacyclic code over  $\mathcal{R}$, where  $\alpha$ is a unit in $\mathcal{R}$ given by \begin{equation}\alpha = \sum_{i,j} \eta_{ij}\alpha_{ij}, ~~\alpha_{ij} \in \mathbb{F}_{p^t}\setminus\{0\},\end{equation} so that $\theta_t(\alpha_{ij})=\alpha_{ij}$ and $\theta_t(\alpha)=\alpha$.\vspace{2mm}

\noindent Note that $\alpha^2=1$ if and only if  $\alpha_{ij}^2=1$, i.e. if and only if $\alpha_{ij} = \pm 1$. \vspace{2mm}

In the special case when $\theta_t=$ identity map, we get all the corresponding results for   $\alpha$-constacyclic codes over  $\mathcal{R}$. We shall call $\theta_t$-skew constacyclic code simply as skew constacyclic code. \vspace{2mm}

 A linear code $\mathcal{C}$ of length $n$  over $\mathcal{R}$ is said to be skew $\alpha$-constacyclic code if  $\mathcal{C}$ is invariant under the skew $\alpha$-constacyclic shift $\vartheta_\alpha$, where $\vartheta_\alpha : \mathcal{R}^n
\rightarrow \mathcal{R}^n$ is defined as \begin{equation}\vartheta_\alpha(c_0,c_1,\cdots,c_{n-1})=\big(\alpha\theta_t(c_{n-1}),\theta_t(c_0),\cdots,\theta_t(c_{n-2})\big),\end{equation} i.e.,
$\mathcal{C}$ is  skew $\alpha$-constacyclic code if and only if $\vartheta_\alpha(\mathcal{C})=\mathcal{C}$. Clearly $\mathcal{C}$  is  skew cyclic  if $\alpha=1$ and is called skew negacyclic  if $\alpha=-1$. \vspace{2mm}

By identifying each codeword by the corresponding polynomial, a linear code $\mathcal{C}$ of length $n$  over $\mathcal{R}$ is  skew $\alpha$-constacyclic code if and only if is left  $\mathcal{R}[x,\theta_t]$-submodule of left  $\mathcal{R}[x,\theta_t]$-module $\mathcal{R}_{n,\alpha}= \mathcal{R}[x,\theta_t]/\langle x^n-\alpha\rangle$.

\begin{theorem} Let the unit $\alpha$ be as defined in (8). A linear code  $\mathcal{C}=\underset{i,j}{\bigoplus}~\eta_{ij}\mathcal{C}_{ij}$ is a skew $\alpha$-constacyclic code of length $n$  over $\mathcal{R}$ if and only if $\mathcal{C}_{ij}$ are skew $\alpha_{ij}$-constacyclic code of length $n$  over $\mathbb{F}_{q}$.\end{theorem}

\noindent{\bf Proof:} Let $c=(c_0,c_1,\cdots,c_{n-1})\in \mathcal{C}$, where
$c_s=\sum_{i,j}\eta_{ij}a_{ij}^{(s)}$ for each $s, 0\le s \le n-1$. Let $a_{ij}= ( a_{ij}^{(0)},a_{ij}^{(1)},\cdots,a_{ij}^{(n-1)})$ so that $c=\sum_{i,j}\eta_{ij}a_{ij}$, $ a_{ij}\in \mathcal{C}_{ij}$. Note that, using the properties of idempotents $\eta_{ij}$ from Lemma 1 $$ \alpha c_{n-1}= \Big(\sum_{i,j}\eta_{ij}\alpha_{ij}\Big) \Big(\sum_{i,j}\eta_{ij}a_{ij}^{(n-1)}\Big)= \sum_{i,j}\eta_{ij}\alpha_{ij}a_{ij}^{(n-1)}.$$

Therefore $$\begin{array}{ll} \vartheta_{\alpha}(c)&= (\alpha \theta_t(c_{n-1}), \theta_t(c_{0}),\cdots,\theta_t(c_{n-2})\vspace{2mm}\\&= (\theta_t(\alpha c_{n-1}), \theta_t(c_{0}),\cdots,\theta_t(c_{n-2})\vspace{2mm}\\&= \Big(\theta_t(\sum_{i,j}\eta_{ij}\alpha_{ij}a_{ij}^{(n-1)}), \theta_t(\sum_{i,j}\eta_{ij}a_{ij}^{(0)}),\cdots,\theta_t(\sum_{i,j}\eta_{ij}a_{ij}^{(n-2)})\Big)\vspace{2mm}\\&= \Big(\sum_{i,j}\eta_{ij}\alpha_{ij}\theta_t(a_{ij}^{(n-1)}), \sum_{i,j}\eta_{ij}\theta_t(a_{ij}^{(0)}),\cdots,\sum_{i,j}\eta_{ij}\theta_t(a_{ij}^{(n-2)})\Big)\vspace{2mm}\\&= \sum_{i,j}\eta_{ij}\Big(\alpha_{ij}\theta_t(a_{ij}^{(n-1)}), \theta_t(a_{ij}^{(0)}),\cdots,\theta_t(a_{ij}^{(n-2)})\Big)\vspace{2mm}\\&=\sum_{i,j}\eta_{ij}\vartheta_{\alpha_{ij}}(a_{ij}).\end{array}$$

Therefore $\vartheta_{\alpha}(c) \in \mathcal{C} $ if and only if $\vartheta_{\alpha_{ij}}(a_{ij}) \in \mathcal{C}_{ij}$.  \hfill $\square$\vspace{2mm}

\noindent{\bf Example 4: } Let $f(u)=u^4-u= u(u-1)(u-\xi)(u-\xi^2),$ where $\xi \in \mathbb{F}_q, ~ \xi^3=1$ and $q\equiv 1 (\mod 3)$. Let $g(v)=v$ so that $\eta_{11}=1-u^3$, $\eta_{21}= \frac{1}{3}(u-\xi)(u-\xi^2)$, $\eta_{31}= \frac{1}{3}(u-1)(u-\xi^2)$ and $\eta_{41}= \frac{1}{3}(u-1)(u-\xi)$. Take $\theta_t=$ Identity automorphism and the unit $\alpha=\eta_{11}-\eta_{21}-\eta_{31}-\eta_{41}=1-2u^3$.  Then a linear code $\mathcal{C} $ is $(1-2u^3)$-constacyclic code over $\mathcal{R}=\mathbb{F}_q[u]/\langle u^4-u\rangle$ if and only if $\mathcal{C}_{11}$ is cyclic and $\mathcal{C}_{21}$, $\mathcal{C}_{31}$, $\mathcal{C}_{41}$ are negacyclic codes of length $n$  over $\mathbb{F}_{q}$. This is Theorem 2 of \cite{RKG}.\vspace{2mm}

\noindent Following is Lemma 3.1 of Jitman et al.  \cite{Jit2012}, where $R$ was taken as a finite chain ring, but the result is true for any finite ring. \vspace{2mm}

\noindent {\bf Lemma 2 } Let $C$ be a linear code of length $n$ over a finite ring $R$. Let $\theta$ be an automorphism of $R$ and suppose $n$ is a multiple of the order of $\theta$. Let $\lambda$ be a unit in $R$ such that $\theta(\lambda)=\lambda$. Then $C$ is skew $\lambda$-constacyclic code if and only if $C^\perp$ is skew $\lambda^{-1}$-constacyclic code over $R$.

\begin{theorem} Let the order of $\theta_t$ divide $n$. If the code  $\mathcal{C}=\underset{i,j}{\bigoplus}~\eta_{ij}\mathcal{C}_{ij}$ is a skew $\alpha$-constacyclic  of length $n$  over $\mathcal{R}$, then $\mathcal{C}^\perp$ is skew $\alpha^{-1}$-constacyclic code over $\mathcal{R}$ and $\mathcal{C}_{ij}^\perp$ are $\alpha_{ij}^{-1}$-constacyclic codes over $\mathbb{F}_{q}$, where $\alpha$ is as given in (8).  Further for $\mathcal{C}$ to be self-dual  it is necessary that $\alpha=  \sum_{i,j} (\pm\eta_{ij})$, i.e. $\alpha^2=1$.\end{theorem}

\noindent{\bf Proof:} The first statement follows from Lemma 2, as $\theta_t(\alpha)=\alpha$. Also by Theorem 1, we have $\mathcal{C}^\perp=\underset{i,j}{\bigoplus}~\eta_{ij}\mathcal{C}_{ij}^\perp$, and $\alpha^{-1} = \sum_{i,j} \eta_{ij}\alpha_{ij}^{-1}$. Therefore $\mathcal{C}_{ij}^\perp$ are $\alpha_{ij}^{-1}$-constacyclic codes over $\mathbb{F}_{q}$. Further $\mathcal{C}$ is self-dual if and only if $\mathcal{C}_{ij}$ are self-dual. Now for $\mathcal{C}_{ij}$ to be self dual  it is necessary that  $\alpha_{ij}=\alpha_{ij}^{-1}$ in $\mathbb{F}_{q}$ i.e.  $\alpha_{ij}=\pm 1$. \hfill $\square$\vspace{2mm}

\noindent{\bf Remark 3:} It may happen that $\alpha= \sum_{i,j}(\pm \eta_{ij})$, i.e. $\alpha_{ij}=\pm 1$, but $\mathcal{C}_{ij}$ are not self-dual skew $\alpha_{ij}$-constacyclic codes and so $\mathcal{C}$ may not be  self-dual skew $\alpha$-constacyclic code.

\begin{cor} Let the order of $\theta_t$ divide $n$. Then the number of units $\alpha$ for which $\mathcal{C}$ can be self-dual skew  $\alpha$-constacyclic  of length $n$  over $\mathcal{R}$
 is $2^{k\ell}$. \end{cor}

Gao et al.\cite{Gao} showed that a skew $\lambda$-constacyclic code  of length $n$ over $\mathbb{F}_q$  is generated by a monic polynomial $g(x)$ which is a right divisor of $x^n-\lambda$ in $\mathbb{F}_q[x;\theta_t]$.  Analogous to this we have the following results for skew constacyclic  codes over $\mathcal{R}$.

\begin{theorem}  Let  $\mathcal{C}=\underset{i,j}{\bigoplus}~\eta_{ij}\mathcal{C}_{ij}$ be a skew $\alpha$-constacyclic code  of length $n$ over $\mathcal{R}$.   Suppose that skew $\alpha_{ij}$-constacyclic codes  $\mathcal{C}_{ij}=\langle g_{ij}(x)\rangle $, where $g_{ij}(x)$ are right divisors of $x^n-\alpha_{ij}$ for $ ~1 \leq i \leq k, 1 \leq j \leq \ell$. Then there exists a polynomial $g(x)$ in $\mathcal{R}[x,\theta_t]$ such that \vspace{2mm}

\noindent (i) $\mathcal{C}=\langle g(x)\rangle$ \vspace{2mm}

\noindent (ii)  $g(x)$ is a right divisor of $(x^{n}-\alpha)$.\vspace{2mm}

\noindent (iii)~~  $|\mathcal{C }|=q^{k\ell n-\sum_{j=1}^{\ell} \sum_{i=1}^{k}deg(g_{ij})}$.\vspace{2mm}
	
\end{theorem}
\noindent{\bf Proof:} First we show that $\mathcal{C}=\langle \eta_{11}g_{11}(x),\cdots,\eta_{1\ell}g_{1\ell}(x),\eta_{21}g_{21}(x),\cdots,\eta_{2\ell}g_{2\ell}(x),$ $\cdots,\eta_{k1}g_{k1}(x),\cdots,
\eta_{k\ell}g_{k\ell}(x)\rangle=\mathcal{E}$, say. \\Let $c(x) \in \mathcal{C}$. Since $\mathcal{C}_{ij}=\langle g_{ij} \rangle$ and $\mathcal{C}=\underset{i,j}{\bigoplus}~\eta_{ij}\mathcal{C}_{ij}$, we have  $c(x)=\underset{i,j}{\sum}~\eta_{ij}u_{ij}(x)g_{ij}(x)$ for $u_{ij}(x) \in
\mathbb{F}_q[x;\theta_t]$. Therefore $c(x)\in \mathcal{E}$ and so $\mathcal{C} \subseteq \mathcal{E}$. \vspace{2mm}

 Conversely let $c(x)=\underset{i,j}{\sum}~\eta_{ij}f_{ij}(x)g_{ij}(x) \in \mathcal{E}$, where $f_{ij}(x) \in
\mathcal{R}[x;\theta_t]$. As $\mathcal{R}= \underset{r,s}{\bigoplus}~\eta_{rs}\mathbb{F}_q$, each $f_{ij}(x)= \underset{r,s}{\sum}~\eta_{rs}u_{rs}(x)$ for some $u_{rs}(x) \in
\mathbb{F}_q[x;\theta_t]$. Now $\eta_{ij}f_{ij}(x)=\eta_{ij}u_{ij}(x)$ as $\eta_{ij}$ are primitive orthogonal idempotents, we see find that $c(x)=\underset{i,j}{\sum}~\eta_{ij}u_{ij}(x)g_{ij}(x)\in \underset{i,j}{\bigoplus}~\eta_{ij}\langle g_{ij}(x)\rangle = \mathcal{C}$, hence $ \mathcal{C} = \mathcal{E}$. \vspace{2mm}

Let $g(x)= \sum_{i}\sum_{j} \eta_{ij}g_{ij}(x)$. Then clearly $\langle g(x) \rangle \subseteq \mathcal{E}=\mathcal{C}$. On the other hand  $\eta_{ij}g(x)=\eta_{ij}g_{ij}(x)$, so $\mathcal{C}\subseteq \langle g(x) \rangle$. \vspace{2mm}

Let for $1\leq i\leq k, 1\leq j\leq \ell$, $x^n-\alpha_{ij}=h_{ij}(x)\ast g_{ij}(x)$ for some $h_{ij}(x)\in \mathbb{F}_q[x;\theta_t]$. Let $ h(x)= \underset{i,j}{\sum}~\eta_{ij}h_{ij}(x),$ then one finds that
		$h(x)\ast g(x)=x^n-\alpha$ so $g(x)$ is a right divisor of $x^n-\alpha$.\vspace{2mm}

Since $|\mathcal{C }|= \underset{i,j} {\prod}|\mathcal{C }_{ij}|$ and $|\mathcal{C }_{ij}|=q^{ n-deg(g_{ij})}$ we get (iii). \hfill $\square$\vspace{2mm}

Next we determine generator polynomial of dual of a skew $\alpha$-constacyclic  codes over $\mathcal{R}$, when  the order of $\theta_t$ divide $n$. First we have
\vspace{2mm}

\noindent{\bf Lemma 3:} Let  order of $\theta_t$ divide $n$ and $ \mathcal{D}=\langle g(x)\rangle$ be a skew $\lambda$-constacyclic code of length $n$ over $\mathbb{F}_q$ then the dual code $ \mathcal{D}^\perp$ is  a skew $\lambda^{-1}$-constacyclic code generated by $h^\perp(x)= h_{n-r}+\theta_t(h_{n-r-1})x+\cdots +\theta_t^{n-r}(h_0)x^{n-r}$, where  $x^n-\lambda=h(x)\ast g(x)$ and  $h(x)=\sum_{i=0}^{n-r}h_ix^i$.  \vspace{2mm}

 \noindent The proof is similar to that of  Corollary 18 of Boucher et al. \cite{BU2009}, where generator of dual of  a  skew cyclic code was determined.

\begin{theorem} Let the order of $\theta_t$ divide $n$. Let  $\mathcal{C}=\underset{i,j}{\bigoplus}~\eta_{ij}\mathcal{C}_{ij}$ be a skew $\alpha$-constacyclic code  of length $n$ over $\mathcal{R}$.   Suppose $\mathcal{C}_{ij}=\langle g_{ij}(x)\rangle $, where  $x^n-\alpha_{ij}=h_{ij}(x)\ast g_{ij}(x)$ for $ ~1 \leq i \leq k, 1 \leq j \leq \ell$. Then \vspace{2mm}

\noindent (i) ~~$ \mathcal{C}^\perp=\langle h^\perp(x)\rangle,$ where
		 $ h^\perp(x)=\sum_{i}\sum_{j}\eta_{ij}h_{ij}^\perp(x)$,
		 \vspace{2mm}
		
		\noindent(ii)$~~ |\mathcal{C}^\perp|=q^{\sum_{j=1}^\ell\sum_{i=1}^k deg(g_{ij})}$.
\end{theorem}

\noindent{\bf Proof :}  By Theorem 1, we have $ \mathcal{C}^\perp=\underset{i,j}{\bigoplus}~\eta_{ij}\mathcal{C}_{ij}^\perp.$ Also, by Lemma 3, $\mathcal{C}_{ij}^\perp= \langle h_{ij}^\perp(x) \rangle$, we get $\mathcal{C}^\perp= \langle h^\perp(x) \rangle$, where $h^\perp(x)= \sum_{i}\sum_{j} \eta_{ij}h_{ij}^\perp$. (ii) follows because $|\mathcal{C }||\mathcal{C^\perp }|= q^{k\ell n}$. \hfill $\square$\vspace{2mm}

Next we compute idempotent generator of the skew constacyclic code $\mathcal{C}$ over $\mathcal{R}$. First we have \vspace{2mm}
		
\noindent{\bf Lemma 4} Let $\mathcal{D}$ be  a $\theta_t$-skew $\lambda$-constacyclic code  of length $n$ over $\mathbb{F}_q$.  If $gcd(n,q)=1$ and $gcd(n,|\theta_t|)=1$, then there exists an idempotent polynomial $e(x)\in \mathbb{F}_{q}[x,\theta_t]/\langle x^n-\lambda \rangle$ such that $\mathcal{D}=\langle e(x)\rangle$. \vspace{2mm}

\noindent The proof is similar to that of Theorem 6 of Gursoy et al. \cite{Gur2014}.

\begin{theorem}  Let  $\mathcal{C}=\underset{i,j}{\bigoplus}~\eta_{ij}\mathcal{C}_{ij}$ be a $\theta_t$-skew $\alpha$-constacyclic code of length $n$ over $\mathcal{R}$. If $gcd(n,q)=1$ and $gcd(n,|\theta_t|)=1$, then there exists an idempotent polynomial $e(x) \in \mathcal{R}[x,\theta_t]/\langle x^n-\alpha \rangle$ such that $\mathcal{C}=\langle e(x)\rangle$ and $\mathcal{C}^\perp=\langle 1- e(x^{-1})\rangle$. \end{theorem}

\noindent{\bf Proof:} By Lemma 4, let $e_{ij}(x) \in \mathbb{F}_q[x,\theta_t]/\langle x^n-\alpha_{ij} \rangle$ be idempotent generators of skew $\alpha_{ij}$-constacyclic codes $\mathcal{C}_{ij}$. Take $e(x)= \sum_{i}\sum_{j} \eta_{ij}e_{ij}(x)$. Then $e(x)$ is an idempotent and also a generator of  $\mathcal{C}$.\vspace{2mm}\\ As $\mathcal{C}_{ij}^\perp$ have idempotent generators $1- e_{ij}(x^{-1})$, $\mathcal{C}^\perp$ has idempotent generator $\sum_{i}\sum_{j} \eta_{ij}\big(1-e_{ij}(x^{-1})\big)= (1-e(x^{-1}))$. \hfill $\square$\vspace{3mm}

Let $c= (c_0,c_1,\cdots,c_{n-1}) = \big(c^{(0)}|c^{(1)}|\cdots |c^{(m-1)}\big)$ be a vector in  $\mathcal{R}^n$ divided into $m$ equal parts of length $r$ where $n=mr$.  We define two skew  $\alpha$-quasi twisted shifts $\varrho_{\alpha, m}$ and $\rho_{\alpha,m}$  as
\begin{equation}\varrho_{\alpha,m}(c)=\big(\alpha \theta_t(c^{(m-1)})|\theta_t(c^{(0)})|\cdots |\theta_t(c^{(m-2)})\big).\end{equation}
\begin{equation}\rho_{\alpha, m}(c)=\big(\vartheta_{\alpha}(c^{(0)})|\vartheta_{\alpha}(c^{(1)})|\cdots |\vartheta_{\alpha}(c^{(m-1)})\big),\end{equation}
where $\vartheta_{\alpha}$ is as defined in (9).

\noindent A linear code $C$  of length $n$ over $\mathcal{R}$ is called  a skew $\alpha$-quasi twisted code of index $m$ if $\varrho_{\alpha,m}(C)=C$ or   $\rho_{\alpha, m}(C)=C$.

\begin{theorem} Let   $\mathcal{C}$ be a skew $\alpha$-constacyclic code of length $n$ over $\mathcal{R}$ and let $r=\gcd (n,|\theta_{t}|)$. If $r=1$, then $\mathcal{C}$ is $\alpha$-constacyclic code of length $n$ over $\mathcal{R}$; If $r> 1$ then $\mathcal{C}$ is a $\alpha$- quasi-twisted code of index $n/r$.\end{theorem}

\noindent{\bf Proof:} Let $n=mr$. Find integers $a$ and $b>0$ such that $ a |\theta_{t}|= r+bn$.  Let $c= (c_0,c_1,\cdots,c_{n-1}) = \big(c^{(0)}|c^{(1)}|\cdots |c^{(m-1)}\big)$ be a codeword in $\mathcal{C}$ divided into $m$ equal parts of length $r$. Since $\mathcal{C}$ is a skew $\alpha$-constacyclic code, $\vartheta_{\alpha}(c), \vartheta_{\alpha}^2(c),\cdots, \vartheta_{\alpha}^r(c),\cdots $ all belong to
$\mathcal{C}$.  Now $$ \begin{array}{rl}~~~~~\vartheta_{\alpha}^r(c)&= \big(\alpha \theta_t^r(c^{(m-1)})|\theta_t^r(c^{(0)})|\cdots |\theta_t^r(c^{(m-2)})\big) \vspace{2mm}\\
~~~~~ \vartheta_{\alpha}^{2r}(c)&= \big(\alpha \theta_t^{2r}(c^{(m-1)})|\alpha\theta_t^{2r}(c^{(0)})|\cdots |\theta_t^{2r}(c^{(m-2)})\big) \vspace{2mm}\\
~~~~~ \vartheta_{\alpha}^n(c)&= \big(\alpha \theta_t^n(c^{(0)})|\alpha\theta_t^n(c^{(1)})|\cdots |\alpha\theta_t^n(c^{(m-1)})\big) \vspace{2mm}\\
 ~~~~\vartheta_{\alpha}^{bn}(c)&= \big(\alpha^b \theta_t^{bn}(c^{(0)})|\alpha^b\theta_t^{bn}(c^{(1)})|\cdots |\alpha^b\theta_t^{bn}(c^{(m-1)})\big) \vspace{2mm}\\
\vartheta_{\alpha}^{r+bn}(c)&= \big(\alpha^{b+1} \theta_t^{r+bn}(c^{(m-1)})|\alpha^b\theta_t^{r+bn}(c^{(0)})|\cdots |\alpha^b\theta_t^{r+bn}(c^{(m-2)})\big)\vspace{2mm}\\&=\big(\alpha^{b+1} c^{(m-1)}|\alpha^bc^{(0)}|\cdots |\alpha^bc^{(m-2)}\big)\vspace{2mm}\\&=\alpha^b\big(\alpha c^{(m-1)}|c^{(0)}|\cdots |c^{(m-2)}\big)=\alpha^b \varrho_{\alpha,m}(c),\end{array}$$

\noindent as $r+bn$ is a multiple of order of $\theta_t$.   Therefore $\varrho_{\alpha,m}(c) \in \mathcal{C}$, with $\theta_t=$Identity automorphism. If $r>1$, $\mathcal{C}$ is a $\alpha$- quasi-twisted code of index $m$. If $r=1$, i.e. $m=n$, then
$\mathcal{C}$ is $\alpha$-constacyclic code of length $n$ over $\mathcal{R}$.\hfill $\square$\vspace{2mm}

\noindent {\bf Example 5 } Consider the field $\mathbb{F}_9=\mathbb{F}_3[\beta]$, where $ \beta^2+\beta-1=0$ and $\theta$ be the Frobenius automorphism on $\mathbb{F}_9$ defined by $\theta(\beta)=\beta^3$. Let $f(u)=u^3-u$, $g(v)=v^2-1$ and $\mathcal{R}=\mathbb{F}_9[u,v]/\langle u^3-u,v^2-1,uv-vu\rangle$. Take $\alpha =1-u^2-u^2v$ a unit in $\mathcal{R}$. The polynomial $h(x)=x^6+\alpha x^5+x^4+\alpha x^3+x^2+\alpha x+1$ is a right divisor of $x^7-\alpha$ in $\mathcal{R}[x,\theta]$. Also $\gcd(n,|\theta|)=\gcd(7,2)=1$. Therefore $\mathcal{C}=\langle h(x)\rangle$ is a $(1-u^2-u^2v)$-constacyclic code of length 7 over $\mathcal{R}$. \vspace{2mm}

In fact if $\alpha $  is any unit in $\mathcal{R}=\mathbb{F}_{p^2}[u,v]/\langle f(u),g(v),uv-vu\rangle$ satisfying $\alpha^2=1$, i.e. $\alpha= \sum_{i,j} (\pm\eta_{ij})$ and $n$ is odd then $x^n-\alpha=(x-\alpha)(x^{n-1}+\alpha x^{n-2}+x^{n-3}+\cdots+\alpha x^3+x^2+\alpha x+1)$. Therefore the skew $\alpha$-constacyclic code  $\mathcal{C}=\langle x^{n-1}+\alpha x^{n-2}+x^{n-3}+\cdots+\alpha x^3+x^2+\alpha x+1\rangle$ is a $\alpha$-constacyclic code of length $n$ over $\mathcal{R}$. \vspace{2mm}

\noindent {\bf Example 6 } Consider the field $\mathbb{F}_{25}=\mathbb{F}_5[\beta]$, where $ \beta^2-\beta+2=0$ and $\theta$ be the Frobenius automorphism on $\mathbb{F}_{25}$ defined by $\theta(\beta)=\beta^5$. Let $f(u)=u^3-u$, $g(v)=v^2-v$ and $\mathcal{R}=\mathbb{F}_{25}[u,v]/\langle u^3-u,v^2-1,uv-vu\rangle$. Now $x^6-1=(x^2-1)(x^2-x+1)(x^2+x+1)$ and $x^6+1= (x^2+1)(x^2+2x-1)(x^2+3x-1)$. Let $\alpha=\eta_{11}+\eta_{12}-\eta_{21}+\eta_{22}-\eta_{31}+\eta_{32}= 1-2u^2+2vu^2$, $g_{11}= g_{12}=g_{22}=g_{32}=x^2+x+1$ and $g_{21}=g_{31}=x^2+3x-1$. Then $\mathcal{C}=\langle g(x)\rangle$, where $g(x)= \sum_{i}\sum_{j} \eta_{ij}g_{ij}=x^2+(1+2u^2-2u^2v)x+(1-2u^2+2u^2v)$ is a skew $(1-2u^2+2vu^2)$-constacyclic code of length 6 over $\mathcal{R}$. Further as $\gcd (6,|\theta|)=2$, $\mathcal{C}$ is a $(1-2u^2+2vu^2)$- quasi-twisted code of index $3$.

\begin{theorem} Let $\vartheta_{\alpha}$ be the skew $\alpha$-constacyclic shift defined in (9), $\rho_{\alpha, k\ell}$ be the  $\alpha$-quasi twisted shift as defined in (11) and let $\Phi_\pi$ be the Gray map as defined in (4). Then $\Phi_\pi \sigma_{\alpha}= \rho_{\alpha, k\ell}\Phi_\pi$. \end{theorem}

\noindent {\bf Proof :} Let $r=(r_0,r_1,\cdots,r_{n-1})\in \mathcal{R}^n$, where
$r_s=\eta_{11}a_{11}^{(s)}+\eta_{12}a_{12}^{(s)}+\cdots+\eta_{kl} a_{k\ell}^{(s)}~$. Then $$ \begin{array}{ll}\vartheta_{\alpha}(r)&=\big( \alpha\theta_t(r_{n-1}),\theta_t(r_{0}), \cdots, \theta_t(r_{n-2})\big)\\&= \Big(\alpha \sum_{ij}\eta_{ij}\theta_t(a_{ij}^{(n-1)}), \sum_{ij}\eta_{ij}\theta_t(a_{ij}^{(0)}),\cdots, \sum_{ij}\eta_{ij}\theta_t(a_{ij}^{(n-2)})\Big)\end{array}.$$ Applying $\Phi_\pi$, we get $$\begin{array}{ll}\Phi_\pi(\vartheta_{\alpha}(r))= &\Big(\alpha \theta_t(a_{11}^{(n-1)}),\theta_t(a_{11}^{(0)}), \cdots, \theta_t(a_{11}^{(n-2)}),
\alpha\theta_t(a_{12}^{(n-1)}),\theta_t(a_{12}^{(0)}), \cdots, \theta_t(a_{12}^{(n-2)}),\\& \cdots , \alpha\theta_t(a_{k\ell}^{(n-1)}),\theta_t(a_{k\ell}^{(0)}), \cdots, \theta_t(a_{k\ell}^{(n-2)})\Big)\end{array}$$

\noindent On the other hand $$ \begin{array}{ll}\Phi_\pi(r_0,r_1,\cdots,r_{n-1}) = & \Big(a_{11}^{(0)},a_{11}^{(1)},\cdots,a_{11}^{(n-1)}|a_{12}^{(0)},a_{12}^{(1)},\cdots,a_{12}^{(n-1)}|\\& \cdots| a_{k\ell}^{(0)}, a_{k\ell}^{(1)},\cdots, a_{k\ell}^{(n-1)}\Big).\end{array}$$
 Therefore $$ \begin{array}{ll}\rho_{\alpha,k\ell}(\Phi_\pi(r))= &\Big( \alpha\theta_t(a_{11}^{(n-1)}),\theta_t(a_{11}^{(0)}), \cdots, \theta_t(a_{11}^{(n-2)})|\alpha
\theta_t(a_{12}^{(n-1)}),\theta_t(a_{12}^{(0)}), \cdots, \theta_t(a_{12}^{(n-2)})|\\&\cdots |\alpha\theta_t(a_{k\ell}^{(n-1)}),\theta_t(a_{k\ell}^{(0)}), \cdots, \theta_t(a_{k\ell}^{(n-2)})\Big)\end{array}$$
Hence $\Phi_\pi \vartheta_{\alpha} = \rho_{\alpha,k\ell}\Phi_\pi$.\hfill $\square$

\begin{theorem}  A linear code $\mathcal{C}$ of length $n$ over $\mathcal{R}$ is a skew $\alpha$-constacyclic code if and only if $\Phi_\pi(\mathcal{C})$ is a  skew $\alpha$-quasi-twisted code of length $k\ell n$ over $\mathbb{F}_{q}$  of index $k\ell$.\end{theorem}

\noindent{ \bf Proof :} From Theorem 12, we see that
$$\Phi_\pi (\vartheta_{\alpha}(\mathcal{C}))=\Phi_\pi \sigma_{\alpha}(\mathcal{C})= \rho_{\alpha,k\ell}\Phi_\pi(\mathcal{C})=\rho_{\alpha, k\ell}(\Phi_\pi(\mathcal{C})).$$
Therefore $\vartheta_{\alpha}(\mathcal{C})=\mathcal{C}$ if and only if $\Phi_\pi (\mathcal{C})= \rho_{\alpha, k\ell}(\Phi_\pi(\mathcal{C}))$.\hfill $\square$

\begin{cor} If a linear code $\mathcal{C}$ of length $n$ over $\mathcal{R}$ is a skew $\alpha$-constacyclic (a skew cyclic)  then $\Phi(\mathcal{C})$ is equivalent to a  skew $\alpha$-quasi-twisted (a skew quasi-cyclic) code of length $k\ell n$ over $\mathbb{F}_{q}$  of index $k\ell$.\end{cor}

\noindent {\bf Example 7 } Let $f(u)=u^2-u$, $g(v)=v(v-1)(v-\beta)$ be polynomials over $\mathbb{F}_4=\mathbb{F}_2[\beta]$, where $ \beta^2+\beta+1=0$. Take $\mathcal{R}=\eta_{11}\mathbb{F}_4\oplus\eta_{12}\mathbb{F}_4\oplus\eta_{13}\mathbb{F}_4\oplus\eta_{21}\mathbb{F}_4\oplus\eta_{22}\mathbb{F}_4\oplus \eta_{23}\mathbb{F}_4$. Let $\theta$ be the Frobenius automorphism on $\mathbb{F}_4$ defined by $\theta(\beta)=\beta^2$. A decomposition of $x^6-1$ in the skew polynomial ring $\mathbb{F}_4[x,\theta]$ is $$\begin{array}{ll} x^6-1&=(x^2-1)(x^4+x^2+1)\\&=(x^2-\beta)(x^4+\beta x^2+\beta^2)\\&= (x^2-\beta^2)(x^4+\beta^2 x^2+\beta)\\&=(x^3+\beta x^2+\beta^2 x-\beta^2)(x^3+\beta^2 x^2+\beta^2 x+\beta).\end{array}$$
If we take $\mathcal{C}_{ij}=\langle x^3+\beta^2 x^2+\beta^2 x+\beta \rangle $ for $i=1,2$ and $j=1,2,3$, then $\mathcal{C}=\oplus\eta_{ij} \mathcal{C}_{ij}= \langle x^3+\beta^2 x^2+\beta^2 x+\beta \rangle $ is a skew cyclic code over $\mathcal{R}$ of length 6. Its Gray image $\Phi(\mathcal{C})$ is a quasi-cyclic code of index 6 with parameters $[36,18,4]$. \vspace{2mm}

\noindent If we take $\mathcal{C}_{11} =\mathcal{C}_{13}=\langle x^4+x^2+1\rangle$, $\mathcal{C}_{12} =\mathcal{C}_{21}=\langle x^4+\alpha x^2+\alpha^2\rangle$ and
$\mathcal{C}_{22} =\mathcal{C}_{23}=\langle x^4+\beta^2 x^2+\beta \rangle$, then $\mathcal{C}=\oplus \eta_{ij}\mathcal{C}_{ij}= \langle x^4+(\beta u v^2+v^2+\beta v+\beta^2 u+1)x^2+(\beta^2 uv+\beta^2v^2+v+\beta u+1)$ is a skew cyclic code over $\mathcal{R}$.
Its Gray image $\Phi(\mathcal{C})$ is a quasi-cyclic code of index 6 with parameters $[36,12,3]$.

\begin{theorem} If $n$ is odd and the unit $\alpha$ satisfies $\alpha^2=1$, then a skew $\alpha$-constacyclic code of length $n$  over $\mathcal{R}$ is equivalent to a skew cyclic code over $\mathcal{R}$. \end{theorem}

\noindent{\bf Proof:} Define a map $\varphi : \mathcal{R}_{n}= \mathcal{R}[x,\theta_t]/\langle x^n-1\rangle \rightarrow \mathcal{R}_{n,\alpha}= \mathcal{R}[x,\theta_t]/\langle x^n-\alpha\rangle$ by $\varphi(f(x))=f(\alpha x)$. Then $\varphi$ is $\mathcal{R}[x,\theta_t]$-module isomorphism because
$$\begin{array}{ll} ~~~~~~~~~~~~~~~~~~~~~~~f(x)&=g(x) {\rm ~~in~~} \mathcal{R}[x,\theta_t]/\langle x^n-1\rangle\\ \Leftrightarrow ~~~~~~~~~~~~f(x)-g(x)&=h(x)\ast (x^n-1) {\rm ~ for ~ some ~} h(x)\in \mathcal{R}[x,\theta_t]\\\Leftrightarrow ~~~~~~~~f(\alpha x)-g(\alpha x)&=h(\alpha x)\ast (\alpha^nx^n-1)\\&=h(\alpha x)\ast(\alpha x^n-1) {\rm ~ as ~} \alpha^n=\alpha {\rm ~ for ~} n {\rm ~ odd}\\&=\alpha h(\alpha x)\ast ( x^n-\alpha) {\rm ~ as ~} \alpha^2=1 {\rm ~ and~} \theta_t(\alpha)=\alpha\\ \Leftrightarrow ~~~~~~~~~~~~~~~~~~f(\alpha x)&=g(\alpha x) {\rm ~~in~~} \mathcal{R}[x,\theta_t]/\langle x^n-\alpha\rangle.\end{array}$$ This gives the result. \hfill $\square$
\begin{theorem}  Let  $k\ell \equiv 1 ({\rm mod}~ s/t) $. Then for any $r\in\mathcal{R}^n$, $\Phi \sigma_{\theta_t}(r)= \sigma_{\theta_t}^{k\ell}\Phi(r)$.\end{theorem}

\noindent{ \bf Proof :} Let $r=(r_0,r_1,\cdots,r_{n-1})\in \mathcal{R}^n$, where
$r_s=\sum_{ij}\eta_{ij}a_{ij}^{(s)}$. Then $$ \begin{array}{ll}\Phi(\sigma_{\theta_t}(r))&=\Phi\big( \theta_t(r_{n-1}),\theta_t(r_{0}), \cdots, \theta_t(r_{n-2})\big)\vspace{1mm}\\&= \Phi\Big( \sum_{ij}\eta_{ij}\theta_t(a_{ij}^{(n-1)}), \sum_{ij}\eta_{ij}\theta_t(a_{ij}^{(0)}),\cdots, \sum_{ij}\eta_{ij}\theta_t(a_{ij}^{(n-2)})\Big)\vspace{1mm}\\&= \Big( \Phi\big(\sum_{ij}\eta_{ij}\theta_t(a_{ij}^{(n-1)})\big), \Phi\big(\sum_{ij}\eta_{ij}\theta_t(a_{ij}^{(0)})\big),\cdots, \Phi\big(\sum_{ij}\eta_{ij}\theta_t(a_{ij}^{(n-2)})\big)\Big)  \vspace{1mm}\\&= \Big( \theta_t(a_{11}^{(n-1)}),\theta_t(a_{12}^{(n-1)}),\cdots,\theta_t(a_{k\ell}^{(n-1)}),\theta_t(a_{11}^{(0)}),\theta_t(a_{12}^{(0)}) \cdots, \theta_t(a_{k\ell}^{(0)}),\vspace{1mm}\\&~~~ \cdots, \theta_t(a_{11}^{(n-2)}),\theta_t(a_{12}^{(n-2)}),\cdots, \theta_t(a_{k\ell}^{(n-2)})\Big).\end{array}$$
On the other hand,
$$\begin{array}{ll}\sigma_{\theta_t}(\Phi(r))&= \sigma_{\theta_t} \big( \Phi(r_0),\Phi(r_1),\cdots,\Phi(r_{n-1})\big)\vspace{1mm}\\& =\sigma_{\theta_t} \big(a_{11}^{(0)},a_{12}^{(0)},\cdots,a_{k\ell}^{(0)}, a_{11}^{(1)},a_{12}^{(1)},\cdots, a_{k\ell}^{(1)},\cdots, a_{11}^{(n-1)},\cdots, a_{k\ell}^{(n-1)}\Big)\vspace{1mm}
\\&=\big( \theta_t(a_{k\ell}^{(n-1)}),\theta_t(a_{11}^{(0)}),\theta_t(a_{12}^{(0)}),\cdots,\theta_t(a_{k\ell}^{(0)}),\theta_t( a_{11}^{(1)}),\cdots, \theta_t(a_{k\ell}^{(1)}),\vspace{1mm}\\&~~~~~~\cdots\theta_t( a_{11}^{(n-1)}),\cdots,\theta_t( a_{k\ell-1}^{(n-1)})\Big), \vspace{1mm}
\\
\sigma_{\theta_t}^2(\Phi(r))&= \big( \theta_t^2(a_{k\ell-1}^{(n-1)}),
\theta_t^2(a_{k\ell}^{(n-1)}),\theta_t^2(a_{11}^{(0)}),\theta_t^2(a_{12}^{(0)}),\cdots,\theta_t^2(a_{k\ell}^{(0)}),\theta_t^2( a_{11}^{(1)}),\cdots, \vspace{1mm}\\&~~~~~~\theta_t^2(a_{k\ell}^{(1)}),\cdots,\theta_t^2( a_{11}^{(n-1)}),\cdots,\theta_t^2( a_{k\ell-2}^{(n-1)})\Big),\vspace{1mm}\\
& \cdots  ~~~~~~~~~~~~~~~~~~\cdots ~~~~~~~~~~~~~~~~\cdots\\
\sigma_{\theta_t}^{k\ell}(\Phi(r))&= \big( \theta_t^{k\ell}(a_{11}^{(n-1)}), \theta_t^{k\ell}(a_{12}^{(n-1)}),\cdots,\theta_t^{k\ell}(a_{k\ell}^{(n-1)}),\theta_t^{k\ell}(a_{11}^{(0)}),\theta_t^{k\ell}(a_{12}^{(0)}),
\cdots,
\vspace{1mm}\\&~~~~~~\theta_t^{k\ell}(a_{k\ell}^{(0)}),\theta_t^{k\ell}( a_{11}^{(1)}),\cdots, \theta_t^{k\ell}(a_{k\ell}^{(1)}),\cdots,\theta_t^{k\ell}( a_{11}^{(n-2)}),\cdots,\theta_t^{k\ell}( a_{k\ell}^{(n-2)})\Big).\end{array}$$

\noindent Since here $\theta_t^{k\ell}=\theta_t$, we find that $\Phi \sigma_{\theta_t}(r)= \sigma_{\theta_t}^{k\ell}\Phi(r)$. \hfill $\square$
\begin{cor} If $k\ell \equiv 1 ({\rm mod}~ |\theta_t|) $, then $\mathcal{C}$ is a skew cyclic code if and only if $\Phi(\mathcal{C})$ is fixed by $\sigma_{\theta_t}^{k\ell}$ skew cyclic shift.
\end{cor}
\section{Conclusion}
Let $\mathcal{R}=\mathbb{F}_{q}[u,v]/\langle f(u),g(v),uv-vu\rangle$ be a finite  non-chain ring where $f(u)$ and $g(v)$ are two polynomials of degree $k$ and $\ell$ respectively, which split into distinct linear factors over $\mathbb{F}_{q}$. We assume that at least one of $k$ and $\ell$ is $\geq 2$. In this paper,  we define two automorphisms $\psi$ and $\theta_t$ on  $\mathcal{R}$ and discuss $\psi$-skew cyclic and $\theta_t$-skew $\alpha$-constacyclic codes over $\mathcal{R}$, where $\alpha$ is any unit in $\mathcal{R}$ fixed by the automorphism $\theta_t$, in particular when $\alpha^2=1$. We show that a skew $\alpha$-constacyclic code of length $n$ over $\mathcal{R}$  is either an $\alpha$-constacyclic code or  a $\alpha$- quasi-twisted code.
Some structural properties, specially generator polynomials and idempotent generators for  skew constacyclic codes are determined.
 A Gray map is defined from $\mathcal{R}^n \rightarrow \mathbb{F}^{k\ell n}_q$ which preserves  duality.  It is shown that Gray image of  a $\theta_t$-skew $\alpha$-constacyclic code of length $n$ over $\mathcal{R}$  is a $\theta_t$-skew $\alpha$-quasi-twisted code of length $k\ell n$ over $\mathbb{F}_{q}$  of index $k\ell$.  Some examples are also given to illustrate the theory. \vspace{2mm}

 \noindent {\bf Acknowledgements}:  The research of second author is supported by Council of Scientific and Industrial Research (CSIR), India, sanction no. 21(1042)/17/ EMR-II.

	\end{document}